\documentclass[conference]{IEEEtran}
\IEEEoverridecommandlockouts
\usepackage{amsmath,amsfonts}
\usepackage{algorithmic}
\usepackage{algorithm}
\usepackage{array}
\usepackage[caption=false,font=normalsize,labelfont=sf,textfont=sf]{subfig}
\usepackage{textcomp}
\usepackage{stfloats}
\usepackage{url}
\usepackage{verbatim}
\usepackage{graphicx}
\usepackage{cite}
\usepackage{comment}
\usepackage{pifont} 
\usepackage{caption}

\usepackage{multirow} 
\usepackage{booktabs}
\usepackage{caption}
\usepackage[table,xcdraw]{xcolor}
\usepackage{geometry}
\usepackage[utf8]{inputenc}
\usepackage{booktabs}
\usepackage{multirow}
\usepackage{graphicx} 
\usepackage{geometry} 
\usepackage{lipsum}
\usepackage{graphicx}
\usepackage{subcaption}

\usepackage{xspace}

\begin{document}

\title{A Unified Framework for Context-Aware IoT Management and State-of-the-Art IoT Traffic Anomaly Detection
}

\author{
    Daniel Adu Worae\\
    University of Notre Dame\\
    \texttt{dworae@nd.edu}
    \and
    Athar Sheikh\\
    University of Notre Dame\\
    \texttt{asheikh@nd.edu}
    \and
    Spyridon Mastorakis\\
    University of Notre Dame\\
    \texttt{mastorakis@nd.edu}
}


\maketitle

\begin{abstract}
The rapid expansion of Internet of Things (IoT) ecosystems has introduced growing complexities in device management and network security. To address these challenges, we present a unified framework that combines context-driven large language models (LLMs) for IoT administrative tasks with a fine-tuned anomaly detection module for network traffic analysis. The framework streamlines administrative processes such as device management, troubleshooting, and security enforcement by harnessing contextual knowledge from IoT manuals and operational data.
The anomaly detection model achieves state-of-the-art performance in identifying irregularities and threats within IoT traffic, leveraging fine-tuning to deliver exceptional accuracy. Evaluations demonstrate that incorporating relevant contextual information significantly enhances the precision and reliability of LLM-based responses for diverse IoT administrative tasks. Additionally, resource usage metrics—such as execution time, memory consumption, and response efficiency—demonstrate the framework's scalability and suitability for real-world IoT deployments.
\end{abstract}



\section{Introduction}
The Internet of Things (IoT) is projected to surpass 30 billion interconnected devices by 2030\cite{statista_iot}, revolutionizing industries through automation and data-driven decision-making. Despite its transformative potential, this rapid growth has introduced significant administrative challenges. Managing the heterogeneity and interconnectivity of IoT devices requires administrators to address complex tasks such as configuring devices securely, troubleshooting, and maintaining network integrity. The increasing sophistication of cyber threats exacerbates these challenges, demanding intelligent solutions that ensure operational efficiency and robust security.


\begin{figure}[h]
    \centering
    \includegraphics[width=3.23in]{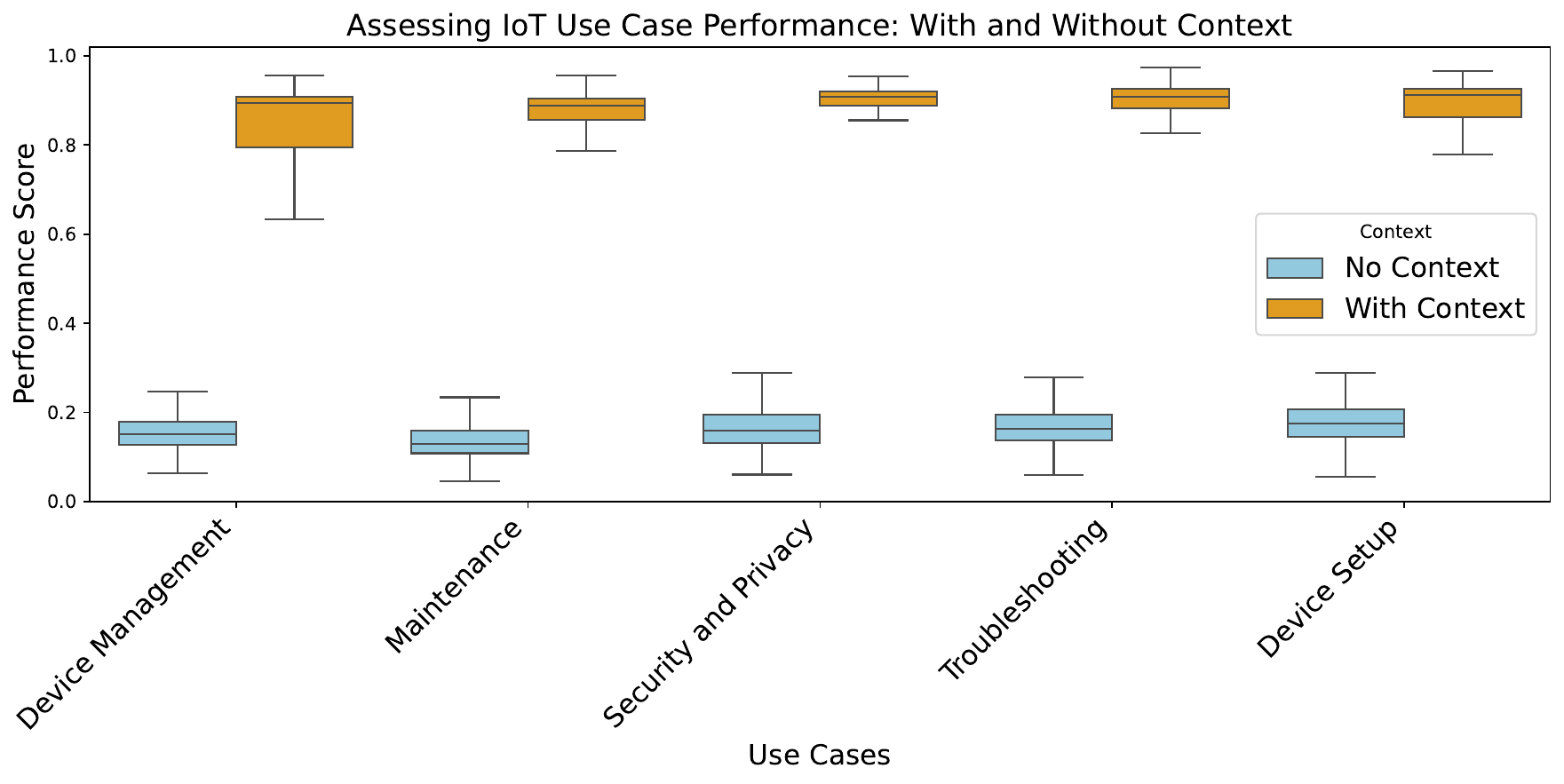}  
    \caption{Comparative performance analysis of IoT use cases with and without contextual augmentation. Grounded in task-specific knowledge retrieval, RAG with LLMs significantly improves performance across all use cases, delivering accurate and context-aware responses. In contrast, the absence of context results in markedly poor performance, highlighting the importance of augmentation.}
    \label{fig1.1}
\end{figure}

LLMs, such as GPT-4, are promising tools for assisting IoT administrators by providing insights and answering queries across a wide range of tasks\cite{veturi2024rag}. However, their application in IoT environments is hindered by critical limitations, particularly their tendency to hallucinate—generating plausible yet incorrect responses—and reliance on potentially outdated training data. These shortcomings can lead to unreliable outcomes in administrative scenarios where precision is critical. For instance, an administrator might need immediate, precise answers to questions like, “What are the recommended security configurations for a newly deployed IoT device?” or “How can I identify and mitigate a coordinated cyberattack targeting multiple IoT endpoints without risking operational downtime?” A hallucinated response in such cases could lead to severe misconfigurations or overlooked threats, with cascading consequences for network operations and security.


RAG provides a promising solution to address these limitations by dynamically integrating external knowledge repositories. RAG grounds LLM outputs in reliable, domain-specific information\cite{borgeaud2022improving,izacard2023atlas,lewis2020retrieval,khandelwal2019generalization,yasunaga2022retrieval}. This approach significantly improves response accuracy and contextual relevance, equipping administrators with dependable insights for managing IoT devices and networks effectively. Figure 1 illustrates RAG’s impact, showing significant performance gains across five IoT use cases with contextual augmentation, while its absence leads to poorer outcomes, emphasizing its critical role.



Anomaly detection is another critical aspect of IoT management, particularly for identifying potential cyber threats or operational anomalies. This requires nuanced analysis to uncover subtle patterns in network traffic indicative of malicious activities or device malfunctions. Transformer-based models like BERT\cite{kenton2019bert}, renowned for their ability to understand contextual relationships, are particularly well-suited for this task. By fine-tuning BERT on IoT-specific datasets, administrators can achieve precise anomaly detection, enabling proactive responses to emerging security threats. This capability is essential for maintaining IoT ecosystem integrity.


This work introduces a first-of-its-kind IoT management framework that unifies RAG-enhanced LLMs for context-aware question answering with a fine-tuned BERT model for anomaly detection. To the best of our knowledge, this is the only framework explicitly designed to address the unique challenges of administrative management in IoT environments, providing actionable insights and precise anomaly detection tailored to the complexities of IoT ecosystems. By bridging these critical gaps, the framework offers a comprehensive solution for administrators managing IoT networks securely and efficiently. 

To advance the understanding and effectiveness of the framework, we explore the following key research questions:

\begin{itemize}
    \item RQ1: How does integrating RAG with LLMs impact the accuracy and reliability of responses for IoT administrative tasks, including device management, security and privacy, troubleshooting, maintenance, and device setup?
    \item RQ2: How effective is a fine-tuned BERT model in detecting nuanced anomalies within IoT traffic, and what critical insights can it provide to preempt cybersecurity threats and operational disruptions?
    \item RQ3: How can a unified framework effectively integrate RAG-based contextual question answering with BERT-based anomaly detection to address the operational and security challenges IoT administrators face?
\end{itemize}

To answer these research questions, we make the following contributions:


\begin{itemize}
    
    \item We introduce a first-of-its-kind IoT management framework that integrates RAG with LLMs to provide accurate, context-aware responses for administrative tasks, including device setup, maintenance, security, and troubleshooting.
    \item The framework incorporates a fine-tuned BERT model for nuanced anomaly detection in IoT traffic, achieving an accuracy of 99.87\% on the Edge-IIoTset dataset.
    \item We evaluate the framework's question-answering system using four advanced LLMs—Gemma2, Llama 3.2, Mistral, and llava—and benchmark its performance with and without contextual augmentation via RAG. Metrics such as BLEU, ROUGE, METEOR, and BERTScore demonstrate that integrating RAG significantly improves the accuracy and relevance of responses, validating the framework’s effectiveness in addressing IoT-specific queries.
    \item We also assess the system's performance in terms of execution time, GPU usage, and memory consumption, comparing results with and without contextual augmentation via RAG. Token metrics, including token count and response size per LLM, are also assessed. This evaluation provides critical insights into the computational efficiency of our framework, demonstrating its ability to deliver accurate, contextually relevant responses while maintaining optimal resource utilization.
\end{itemize}

The rest of our paper is organized as follows: Section II describes the system design of our framework. Section III presents experimental results and an analysis of our framework. In Section IV, we present a literature review of our framework. Finally, we conclude in Section V.

\section{System Design}
This paper proposes a framework that integrates a context-aware generation module and a fine-tuned BERT-based anomaly detection module, as illustrated in Figures 2 and 3. The framework addresses two critical challenges in IoT ecosystems: providing accurate, domain-specific responses to administrative queries and ensuring real-time detection of cyber threats and anomalies in network traffic. The system achieves this through a modular design, where each component operates cohesively to enhance operational efficiency and security. This section outlines the architecture, interactions, and functionality of these components, forming a robust and scalable solution for modern IoT environments.

\begin{figure}[h]
    \centering
    \includegraphics[width=3.45in]{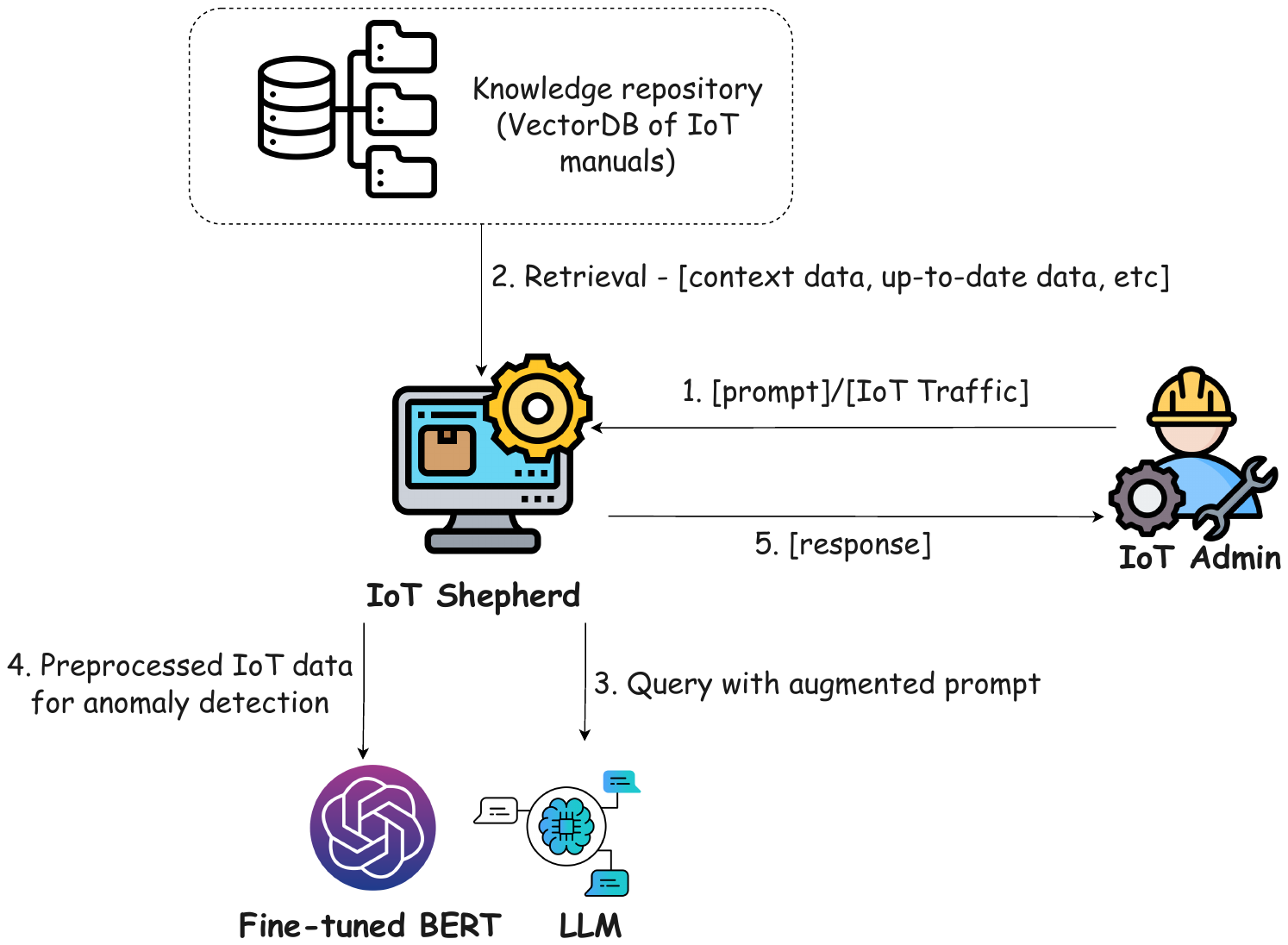}  
    \caption{High-Level Overview of our framework}
    \label{fig1}
\end{figure}

\subsection{Context-Aware Generation Module}
This module operates in two phases: Indexing and Query Processing. During indexing, IoT manuals, documents, and other domain-specific resources are preprocessed, chunked, and stored as vector embeddings in a vector database. During query processing, an administrator’s query is transformed into a vector representation, compared with the stored vectors, and augmented with the most relevant knowledge chunks before being passed to the LLM for final answer generation, as shown in Figure 4. This ensures that the LLM produces domain-specific and accurate responses.

\begin{figure*}[h]
    \centering  \includegraphics[width=0.8\textwidth]{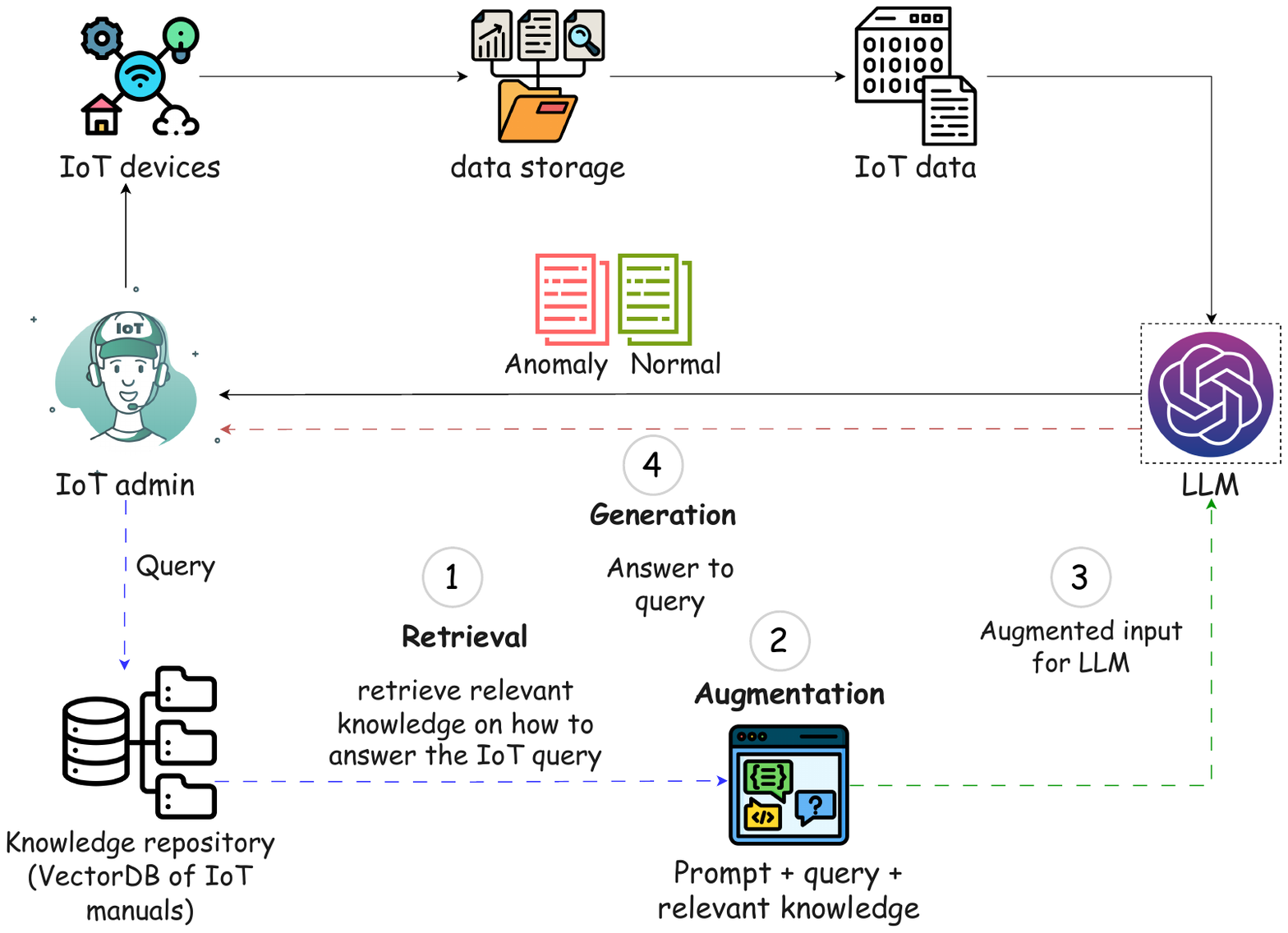} 
    \caption{System Design}
    \label{fig3}
\end{figure*}

\begin{figure*}[h]
    \centering  \includegraphics[width=0.8\textwidth]{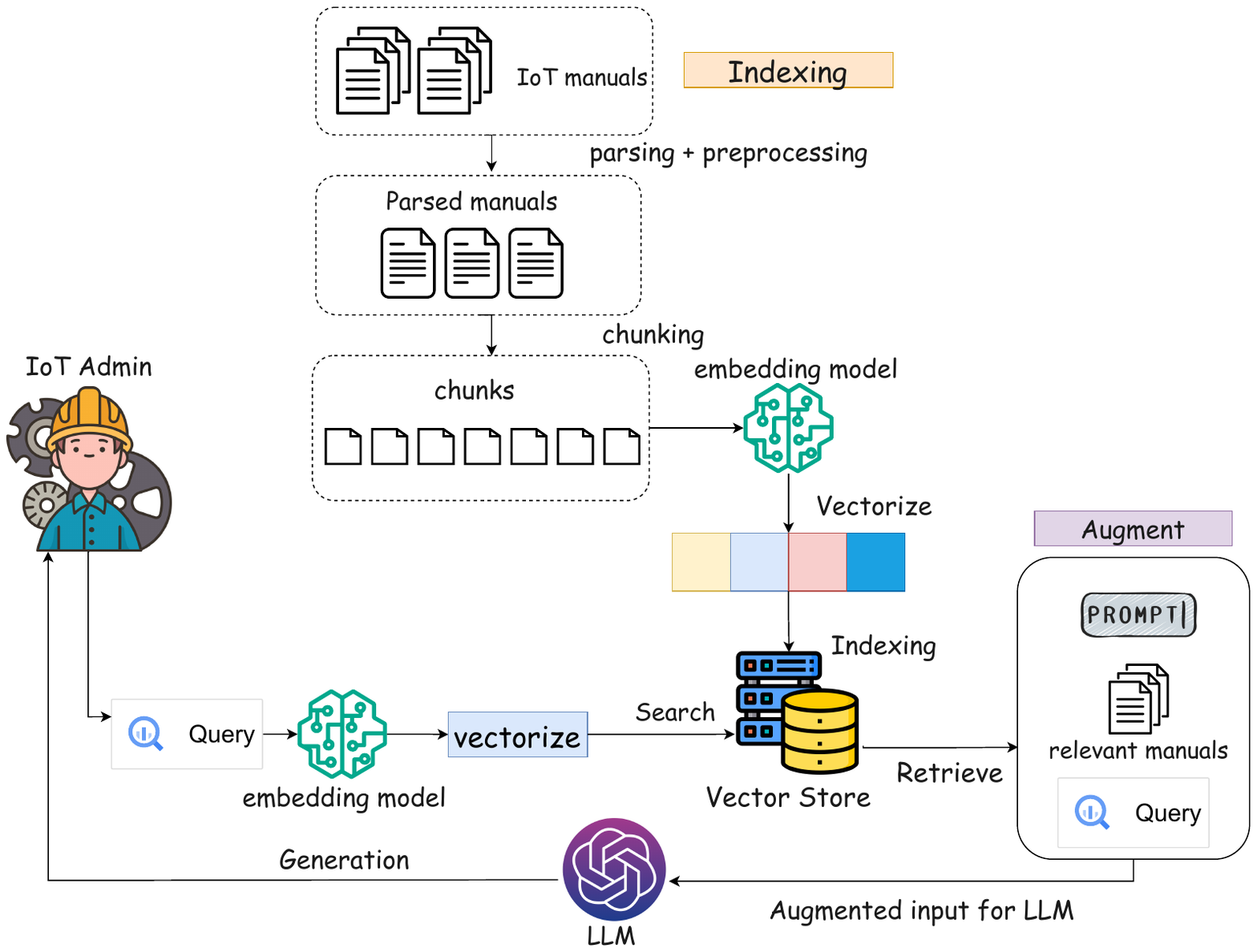} 
    \caption{Retrieval-Augmented Generation Pipeline}
    \label{fig4}
\end{figure*}

\subsubsection{IoT Knowledge Repository}

\begin{itemize}
    \item \textbf{Content Sources} The IoT knowledge repository consists of device manuals, troubleshooting guides, FAQs, and other domain-specific resources. These documents form the backbone of the knowledge hub, ensuring that responses to administrator queries are grounded in accurate, specialized information.
    \item \textbf{Preprocessing:} Documents are parsed and preprocessed to remove irrelevant content (e.g., metadata, unrelated sections). The cleaned documents are then divided into manageable chunks, ensuring each chunk contains semantically meaningful information.
    \item \textbf{Vectorization:} Each chunk is passed through an embedding model (e.g., sentence transformers) to convert into a high-dimensional vector representation. These embeddings capture the semantic meaning of the text and are stored in the vector database. 
\end{itemize}

\subsubsection{Vector Database (Knowledge Hub)}
The vector database stores embeddings of document chunks together with the chunks and enables efficient similarity search for relevant knowledge. In our implementation, we use Chroma to manage these embeddings, leveraging its ability to handle large-scale vectorized data while supporting persistence for future use.

During indexing, IoT manuals are loaded from the data directory and processed using PyPDFDirectoryLoader. Documents are split into smaller, semantically meaningful chunks (800 characters with 80-character overlap) using RecursiveCharacterTextSplitter. Each chunk is converted into an embedding using a custom embedding function, with metadata such as source file, page number, and a unique chunk ID assigned for precise identification.

To prevent duplication, the system compares chunk IDs with existing entries in the database. Only new chunks are added, and the database is persisted to ensure efficient future retrieval. This approach provides scalability and accuracy while maintaining a streamlined knowledge base for IoT-specific query processing.

\subsubsection{Query Processing Workflow} 
\begin{itemize}
    \item \textbf{Query Vectorization:} When the IoT administrator submits a query (e.g., “What steps should I follow to fix Amazon Echo when it cannot control smart locks?”), the query text is passed through the same embedding model used during the indexing phase. This converts the query into a vector representation, ensuring consistency in how text is represented.
    \item \textbf{Similarity Search:} The query embedding is compared to the embeddings in the vector database using a similarity metric (e.g., cosine similarity). This process identifies the chunks of knowledge most semantically similar to the query.
    \item \textbf{Knowledge Retrieval:} The system retrieves the top-ranked chunks from the vector database. These chunks serve as the relevant context for the query, ensuring that the subsequent response generation is grounded in accurate domain-specific knowledge.
    \item \textbf{Context Augmentation:} The retrieved knowledge chunks are concatenated with the user query to form an augmented input. This input is structured as a prompt for the LLM, ensuring the model generates answers based on the query and the retrieved domain-specific context.
\end{itemize}

\subsubsection{Large Language Model} The LLM (e.g., llama or similar model) processes the augmented input and generates a contextually relevant and precise response. By leveraging the retrieved knowledge, the model addresses out-of-date training data, avoids hallucination, and delivers answers tailored to IoT-specific queries. The response is then returned to the IoT administrator for actionable insights.

\subsection{Transformer-Based Anomaly Detection Module}
This module uses a fine-tuned BERT model to detect anomalies in IoT network traffic by identifying complex patterns and contextual relationships. It strengthens IoT security by addressing cyber threats and device malfunctions in real-time, enhancing system reliability.

\subsubsection{\textbf{Dataset Utilization}}
Generating a dataset through real-life IoT traffic analysis is time-intensive and risks omitting critical attack scenarios. To address this, we prioritize the use of realistic and diverse datasets that represent genuine network conditions. The Edge-IIoTset dataset\cite{9751703} provides a comprehensive cybersecurity resource designed specifically for IoT and IIoT applications. It includes fifteen attack types across five major threat categories: DoS/DDoS, Information Gathering, Man-in-the-Middle (MITM), Injection, and Malware. These attacks encompass techniques such as TCP SYN flood, port scanning, DNS spoofing, SQL injection, and ransomware, making the dataset highly representative of real-world IoT threats. Its diversity and focus on IoT connectivity protocols ensure its alignment with our research objectives for anomaly detection and classification.

\subsubsection{\textbf{Feature Selection and Data Preprocessing}}
Network traffic logs in PCAP format are processed to extract features over a defined time window, resulting in structured data suitable for analysis. Each network flow is identified and analyzed to extract a predefined set of features stored in CSV format for efficient processing.
The Edge-IIoTset dataset, after preprocessing to remove null features, provides 61 distinct attributes sourced from network traffic, system logs, resource usage, and alerts. However, not all features are suitable for effective modeling. Features with high cardinalities, such as “http.request.full\_uri”, or those prone to overfitting, like “ip.src\_host” and “ip.dst\_host”, were excluded due to their limited generalizability across networks. Similarly, redundant features such as “arp.src.proto\_ipv4” and “ip.src\_host, or features like “tcp.payload” containing raw data, may introduce noise or multicollinearity, reducing model stability. Removing such columns optimizes the dataset, ensuring computational efficiency while maintaining focus on relevant, generalizable patterns. This refined feature set forms a robust foundation for anomaly detection, tailored to IoT environments and capable of capturing diverse network behaviors effectively.

\subsubsection{\textbf{Contextual Representation}}
The BERT module processes raw network data by converting it into structured, textual representations optimized for natural language processing. This transformation reformats the diverse and often semi-structured features of network traffic logs into a descriptive text format, enabling BERT to harness its advanced contextual understanding capabilities. By encapsulating each data row into a structured, natural-language-like text string, the module effectively captures underlying patterns, relationships, and contextual nuances within network traffic, facilitating more accurate and insightful analysis.

In this approach, every row in the dataset is converted into a single textual string made up of feature-value pairs. Each feature column is assigned a descriptive label, and its corresponding value is appended using a delimiter, such as a colon. For example, a feature representing source IP might appear as ip.src: 192.168.1.1. All feature-value pairs for a row are concatenated into a single text string, creating a comprehensive representation of the network flow.

Columns such as labels or target variables are excluded during this transformation to ensure the focus remains on the input features. The resulting textual representations are stored in a structured format, such as a new column in the dataset, and are used as input for the BERT model. This structured approach allows BERT to analyze network data contextually, capturing meaningful patterns that enable effective anomaly detection and threat identification.

\subsubsection{\textbf{Fine Tuning the BERT module}}
The BERT module customizes a pre-trained BERT model for IoT anomaly detection by training it on labeled network traffic data. This phase builds upon the structured textual representations prepared in the previous stages, aligning them with the model’s requirements for effective learning.

Tokenization is performed using the BERT tokenizer, which converts each textual representation into input embeddings comprising token IDs, attention masks, and segment embeddings. Padding and truncation are applied to ensure uniform input length, enabling the model to process sequences efficiently without exceeding its maximum token limit. The labeled dataset is split into training and testing subsets to support supervised learning and performance evaluation.

The pretrained BERT model is configured for sequence classification by adding a classification head, with output dimensions corresponding to the number of unique attack categories. The model is fine-tuned using the training dataset, leveraging tailored learning parameters, such as batch size, learning rate, and evaluation strategies, to optimize performance. During training, metrics like accuracy, precision, recall, and F1-score are computed to monitor the model’s ability to detect and classify anomalies.

This fine-tuning process equips the BERT model with the capacity to recognize and distinguish complex patterns in IoT network traffic, ensuring robust and scalable anomaly detection.

\subsubsection{\textbf{Testing and Evaluation}}
After fine-tuning, the BERT module is evaluated to validate its performance in detecting and classifying IoT anomalies. The evaluation is conducted using the test dataset, which contains previously unseen network traffic instances to ensure unbiased assessment.

Key metrics such as accuracy, precision, recall, and F1-score are computed to measure the model's effectiveness. Accuracy provides an overall measure of correctness, while precision and recall evaluate the model’s ability to identify anomalies without false positives or missed detections. The F1 score balances these metrics, offering a comprehensive view of the model’s classification capabilities across all attack categories.

The testing process also generates a detailed classification report outlining the model’s performance for each class. This includes insights into attack-specific precision and recall, enabling a granular understanding of how well the model distinguishes among diverse IoT attack types. This thorough testing phase solidifies the model's role as a robust anomaly detection solution capable of addressing the dynamic and complex nature of IoT network security.

\section{Experimental Results and Analysis}
This section evaluates our framework that encompasses RAG-augmented LLMs for context-aware question answering and a fine-tuned BERT model for anomaly detection. The RAG component is analyzed across five IoT use cases to measure the benefits of contextual augmentation. At the same time, the BERT module is tested for its accuracy and robustness in detecting diverse network anomalies, addressing critical challenges in IoT administration and security.

\subsection{Context-Aware Generation Module}
This section evaluates the RAG with LLMs module in addressing IoT-specific queries across five critical use cases: device management, maintenance, security and privacy, troubleshooting, and device setup. To assess the framework's effectiveness, we prepared a dataset of 600 curated question-answer pairs for each use case, representing realistic IoT administrative scenarios derived from device manuals, FAQs, and operational documentation. This evaluation focuses on comparing the performance of LLMs operating in two settings: non-contextual (NC) and contextual (WC), where domain-specific knowledge is dynamically retrieved using the RAG framework.

The analysis, summarized in Table I, is based on four state-of-the-art LLMs—Gemma2, Llama 3.2, Mistral, and Llava—with performance quantified using metrics such as BLEU, ROUGE, METEOR, and BERTScore. These metrics evaluate syntactic accuracy, linguistic quality, and semantic alignment, providing a comprehensive understanding of the framework's impact on IoT administrative tasks.

\subsection*{\textbf{1) Evaluation Metrics}}
To evaluate the generated responses, we employed the following metrics:

\begin{itemize}
    \item {\textit{BERT (Precision/Recall/F1-Score)}:} Measures semantic alignment with ground truth, balancing precision and recall to ensure reliable and robust model behavior.
    \item {\textit{ROUGE (1/2/L)}:} Assesses response completeness and comprehensiveness through overlaps at the unigram (ROUGE-1), bigram (ROUGE-2), and sequence (ROUGE-L) levels. 
    \item {\textit{BLEU}:} Focuses on n-gram precision, highlighting syntactic accuracy and procedural fluency.
    \item {\textit{METEOR}:} Combines precision, recall, and linguistic features such as synonymy and stemming to provide a nuanced evaluation of response quality.
\end{itemize}

\subsection*{\textbf{2) Use Cases and Results}}

\subsubsection*{a) Device Management} This encompasses routine administrative tasks such as firmware updates, performance monitoring, and device health checks. Contextual augmentation yielded substantial improvements in these tasks. Gemma2 (WC) achieved a BLEU score of 70.2, compared to 0.59 (NC), reflecting significant gains in syntactic precision. Similarly, Mistral (WC) recorded a ROUGE-L score of 75.33\%, up from 9.68 (NC), demonstrating enhanced response completeness. BERTScore (WC) for Gemma2 reached 95.8\%, underscoring strong semantic alignment. Additionally, the METEOR score for Llava (WC) increased to 62.63\%, highlighting improved linguistic richness and relevance.

\subsubsection*{b) Maintenance} This involves diagnostic assessments, troubleshooting, and preventive actions to ensure IoT systems remain operational and secure. Contextual augmentation produced notable enhancements in these tasks. Gemma2 (WC) achieved a BLEU score of 71.59, a dramatic improvement over 0.41 (NC). The ROUGE-L score for Mistral (WC) rose to 72.36\%, reflecting increased clarity and comprehensiveness in responses. Additionally, BERTScore (WC) for Mistral reached 93.59\%, compared to 83.24\% (NC), highlighting semantic improvements in diagnostic insights. Llava (WC) achieved a METEOR score of 57.55\%, showcasing improved coherence and linguistic quality.

\subsubsection*{c) Security and privacy} This addresses queries focused on resolving vulnerabilities, mitigating threats, and ensuring compliance with security protocols in IoT ecosystems. These use cases showed the most pronounced gains, underscoring the framework’s value. Gemma2 (WC) achieved a BLEU score of 78.71, far exceeding 0.6 (NC). Mistral (WC) recorded a BERTScore of 95.32\%, compared to 83.88\% (NC), reflecting its ability to generate highly semantically aligned responses. Llava (WC) also delivered a METEOR score of 67.62\%, highlighting its ability to produce linguistically rich and precise outputs for security-focused queries.

\subsubsection*{d) Troubleshooting} These tasks require step-by-step reasoning and detailed explanations to resolve IoT issues efficiently. Contextual augmentation provided significant benefits in this use case. Gemma2 (WC) achieved a BLEU score of 76.11, compared to 0.61 (NC), demonstrating enhanced response fluency. Llama 3.2 (WC) recorded a ROUGE-L score of 68.88\%, reflecting improved structural depth in solutions. Furthermore, Mistral (WC) achieved a BERTScore of 94.24\%, up from 83.71\% (NC), emphasizing its ability to generate coherent and contextually relevant troubleshooting steps.

\subsubsection*{d) Device Setup} This focuses on configuring new devices, integrating them into networks, and ensuring seamless operation through accurate procedural guidance. Contextual augmentation led to significant improvements in these tasks. Gemma2 (WC) achieved a BLEU score of 73.35, a substantial increase from 0.79 (NC). Mistral (WC) demonstrated superior response completeness with a ROUGE-L score of 79.78\% while also achieving a BERTScore of 95.27\%, compared to 84.36\% (NC). Additionally, the METEOR score for Mistral (WC) rose to 75.11\%, underscoring the framework’s ability to deliver clear and accurate procedural guidance.


\begin{table*}[ht]
    \centering
    \captionsetup{justification=centering} 
    \caption{Performance Evaluation of Our Framework's Context-Aware Module for Administrative IoT Tasks}
    \label{tab:evaluation}
    \resizebox{\textwidth}{!}{%
    \begin{tabular}{@{}llcccccccc@{}}
        \toprule
        \textbf{Prompt Type} & \textbf{Metric} 
        & \textbf{gemma2 (NC)} & \textbf{gemma2 (WC)} 
        & \textbf{Llama 3.2 (NC)} & \textbf{Llama 3.2 (WC)} 
        & \textbf{Mistral (NC)} & \textbf{Mistral (WC)} 
        & \textbf{llava (NC)} & \textbf{llava (WC)} \\
        \midrule
        \multirow{4}{*}{\textbf{Device Management}} 
        & BERT (p/r/f) & 78.53 / 87.61 / 82.81 & 96.34 / 95.29 / 95.8 & 78.99 / 87.47 / 83.01 & 92.67 / 94.19 / 93.42 
        & 80.21 / 88.38 / 84.09 & 93.64 / 94.96 / 94.28 & 82.99 / 89.3 / 86.03 & 92.16 / 94.29 / 93.19 \\
        & ROUGE (r1/r2/rL) & 11.03 / 4.96 / 9.08 & 88.23 / 83.67 / 86.32 & 8.64 / 4.01 / 7.53 & 73.46 / 63.6 / 69.92 
        & 11.53 / 5.4 / 9.68 & 75.33 / 67.5 / 72.71 & 16.8 / 7.53 / 13.87 & 66.74 / 55.86 / 62.68 \\
        & BLEU & 0.59 & 70.2 & 0.58 & 39.59 & 0.82 & 47.38 & 1.16 & 33.25 \\
        & METEOR & 15.48 & 76.41 & 14.19 & 66.3 & 17.43 & 71.09 & 21.12 & 62.63 \\
        \midrule
        \multirow{4}{*}{\textbf{Maintenance}} 
        & BERT (p/r/f) & 77.94 / 87.24 / 82.32 & 95.51 / 94.71 / 95.1 & 78.38 / 86.85 / 82.4 & 90.72 / 93.84 / 92.23 
        & 79.42 / 87.47 / 83.24 & 92.65 / 94.58 / 93.59 & 81.46 / 87.88 / 84.54 & 89.56 / 93.58 / 91.51 \\
        & ROUGE (r1/r2/rL) & 8.93 / 3.17 / 6.83 & 86.24 / 82.96 / 85.14 & 6.97 / 2.68 / 5.67 & 66.0 / 57.55 / 64.04 
        & 9.03 / 3.32 / 7.04 & 72.36 / 65.61 / 68.74 & 11.57 / 4.21 / 8.96 & 52.89 / 43.13 / 48.26 \\
        & BLEU & 0.41 & 71.59 & 0.39 & 37.66 & 0.49 & 47.15 & 0.71 & 23.61 \\
        & METEOR & 13.67 & 76.26 & 11.9 & 66.28 & 14.37 & 71.15 & 16.89 & 57.55 \\
        \midrule
        \multirow{4}{*}{\textbf{Security and Privacy}} 
        & BERT (p/r/f) & 78.45 / 87.60 / 82.77 & 97.00 / 95.98 / 96.48 & 79.29 / 87.47 / 83.17 & 94.47 / 95.31 / 94.87 
        & 80.22 / 87.91 / 83.88 & 95.00 / 95.67 / 95.32 & 82.50 / 88.36 / 85.33 & 92.06 / 94.77 / 93.37 \\
        & ROUGE (r1/r2/rL) & 9.88 / 3.62 / 7.44 & 91.81 / 89.21 / 90.79 & 8.52 / 3.44 / 6.81 & 80.79 / 74.19 / 77.90 
        & 10.69 / 4.09 / 8.2 & 81.84 / 75.56 / 78.78 & 14.11 / 5.41 / 10.83 & 64.38 / 55.87 / 60.13 \\
        & BLEU & 0.6 & 78.71 & 0.71 & 57.09 & 0.86 & 59.48 & 1.26 & 38.15 \\
        & METEOR & 16.46 & 82.22 & 15.73 & 76.43 & 18.06 & 78.45 & 21.87 & 67.62 \\
        \midrule
        \multirow{4}{*}{\textbf{Troubleshooting}} 
        & BERT (p/r/f) & 78.26 / 87.35 / 82.55 & 96.58 / 95.53 / 96.05 & 78.87 / 87.06 / 82.76 & 92.06 / 94.66 / 93.32 
        & 79.94 / 87.86 / 83.71 & 93.48 / 95.03 / 94.24 & 82.09 / 88.19 / 85.02 & 91.02 / 94.15 / 92.54 \\
        & ROUGE (r1/r2/rL) & 10.67 / 4.07 / 8.04 & 90.24 / 87.01 / 89.04 & 8.51 / 3.48 / 6.83 & 72.16 / 64.13 / 68.88 
        & 11.35 / 4.6 / 8.73 & 75.74 / 68.65 / 72.41 & 14.69 / 5.54 / 11.12 & 60.8 / 51.03 / 56.45 \\
        & BLEU & 0.61 & 76.11 & 0.56 & 45.20 & 0.8 & 51.38 & 1.02 & 32.12 \\
        & METEOR & 16.78 & 80.64 & 15.04 & 72.65 & 18.20 & 74.84 & 20.71 & 64.19 \\
        \midrule
        \multirow{4}{*}{\textbf{Device Setup}} 
        & BERT (p/r/f) & 78.89 / 87.97 / 83.18 & 96.77 / 95.77 / 96.26 & 79.20 / 87.66 / 83.21 & 94.08 / 95.03 / 94.54 
        & 80.45 / 88.68 / 84.36 & 94.77 / 95.81 / 95.27 & 83.19 / 89.57 / 86.26 & 93.21 / 95.04 / 94.10 \\
        & ROUGE (r1/r2/rL) & 12.65 / 5.6 / 10.18 & 88.11 / 84.26 / 86.57 & 9.81 / 4.65 / 8.36 & 79.62 / 71.15 / 77.11 
        & 13.07 / 6.24 / 11.01 & 81.83 / 74.35 / 79.78 & 18.32 / 8.64 / 15.30 & 72.68 / 62.43 / 69.74 \\
        & BLEU & 0.79 & 73.35 & 0.77 & 51.68 & 1.13 & 57.06 & 1.68 & 43.53 \\
        & METEOR & 17.69 & 78.14 & 16.26 & 71.07 & 19.41 & 75.11 & 23.52 & 66.87 \\
        \bottomrule
    \end{tabular}}
    \caption*{Performance Metrics: p/r/f = Precision/Recall/F1-Score, ROUGE (r1/r2/rL) = ROUGE-1/ROUGE-2/ROUGE-L. \\
    NC = No Context, WC = With Context.}
\end{table*}


\subsection*{\textbf{3) System Performance and Token Metrics Analysis}}
This analysis evaluates execution time, memory usage, GPU utilization, and token metrics, with results averaged across multiple query runs for accuracy and reliability. The findings in Table II provide critical insights into resource efficiency and token utilization across with-context (WC) and without-context (NC) scenarios.

\subsubsection{Execution Time} Context-aware processing significantly improved execution time across all LLMs, exemplified by Gemma2 reducing from 3.8876 seconds (NC) to 0.7095 seconds (WC), a reduction of approximately 81.7\%. This improvement stems from supplying the LLMs with targeted context, which eliminates extra token generation and ensures computational focus on the most relevant information.

\subsubsection{Memory Usage} Memory usage slightly increased when context was added, reflecting the additional processing required for context integration. For example, Gemma2’s memory usage rose from 0.0004 MB (NC) to 0.0295 MB (WC), and similar trends were observed for other models. Despite this increase, memory usage remains negligible for administrative systems, ensuring efficient operation even in resource-constrained environments.

\subsubsection{GPU Usage} GPU utilization remained minimal across all configurations, with the highest value recorded at only 0.0022\% for Llama 3.2 (WC). This demonstrates the lightweight nature of the framework and its suitability for environments with limited GPU resources.

\subsubsection{Token Metrics} Token metrics, including the average number of tokens and response size, show significant reductions in the context-aware setup across all LLMs. Notably, Llama 3.2 processed an average of 440.81 tokens (NC), which dropped to 54.99 tokens (WC), with the corresponding response size decreasing from 2248.55 bytes to 284.99 bytes. These results demonstrate the efficiency of contextual augmentation in streamlining responses and reducing unnecessary generation.

\begin{table*}[ht]
    \centering
    \captionsetup{justification=centering}
    \caption{System Performance and Token Metrics Analysis}
    \label{tab:resource_usage}
    \resizebox{\textwidth}{!}{%
    \begin{tabular}{@{}lcccccccc@{}}
        \toprule
        \textbf{Metric} & \textbf{gemma2 (NC)} & \textbf{gemma2 (WC)} 
        & \textbf{Llama 3.2 (NC)} & \textbf{Llama 3.2 (WC)} 
        & \textbf{Mistral (NC)} & \textbf{Mistral (WC)} 
        & \textbf{llava (NC)} & \textbf{llava (WC)} \\
        \midrule  
        \textbf{Execution Time (s)} & 3.8876 & 0.7095 & 1.8872 & 0.5265 & 2.2351 & 0.7191 & 1.7253 & 0.9250 \\
        \textbf{Memory Usage (MB)} & 0.0004 & 0.0295 & 0.0000 & 0.0287 & 0.0000 & 0.0287 & 0.000 & 0.0287 \\
        \textbf{GPU Utilization (\%)} & 0.0006 & 0.0017 & 0.0010 & 0.0022 & 0.0008 & 0.0021 & 0.0006 & 0.0016 \\
        \textbf{Avg. number of tokens} & 411.7808 & 33.6931 & 440.8145 & 54.9865 & 316.7892 & 54.0287 & 242.1450 & 82.4806 \\
        \textbf{Avg. Response size (bytes)} & 2015.8415 & 174.9578 & 2248.5497 & 284.9899 & 1662.7943 & 294.4637 & 1290.3255 & 450.7487 \\
        \bottomrule
    \end{tabular}}
    \caption*{NC = No Context, WC = With Context.}
\end{table*}

\subsection{Transformer-Based Anomaly Detection Module}
In this section, we assess the performance of our fine-tuned BERT model, which delivers an exceptional accuracy of 99.87\%. This unparalleled result highlights the model’s capacity to accurately identify IoT attacks within complex and realistic network environments, setting a new benchmark for anomaly detection in IoT security.

\subsection*{\textbf{1) Experimental Results}}
We evaluated the model's performance using standard metrics, including Accuracy, Precision, Recall, and F1-Score. The Edge-IIoTset dataset\cite{9751703} was split into 80\% (126,240 data points) for training and 20\% (31,560 data points) for evaluation, ensuring the model was tested on unseen data to validate its anomaly detection capabilities rigorously.

\begin{figure}[h]
    \centering
    \includegraphics[width=3.52in]{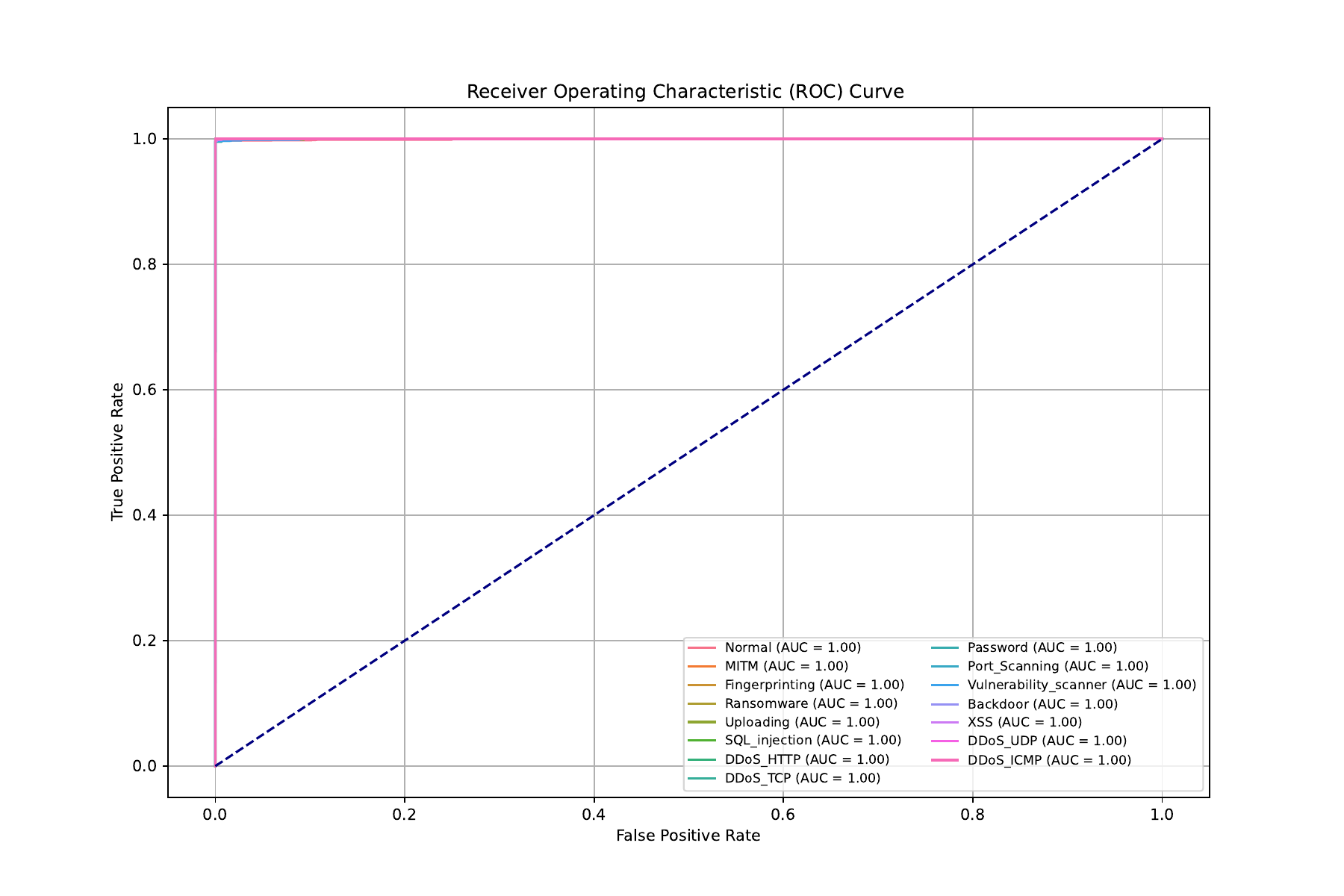}  
    \caption{ROC AUC scores showcasing our fine-tuned BERT model's performance in IoT anomaly detection.}
    \label{fig2}
\end{figure}

\subsection*{a) ROC AUC Scores for Anomaly Detection}
Figure 5 presents the Receiver Operating Characteristic (ROC) curves and corresponding AUC scores for each IoT traffic class, demonstrating the exceptional performance of our fine-tuned BERT model. The model achieves a perfect AUC score of 1.0 for critical classes such as "MITM," "Ransomware," "SQL Injection," "DDoS\_TCP," "Backdoor," "XSS," "DDoS\_HTTP," "Port Scanning," and "Uploading," indicating flawless classification and the ability to distinguish these anomalies from normal traffic with complete accuracy.

For classes like "Normal," "Password," "vulnerability\_scanner," and "DDoS\_ICMP," near-perfect AUC scores ranging from 0.993 to 0.999998 further highlight the model’s precision in detecting nuanced traffic behaviors. These results suggest that the model demonstrates reliable detection with minimal misclassification, even in cases with subtle differences between attack patterns.

The uniform distribution of high AUC scores across all attack vectors validates the model's robustness in identifying diverse IoT threats. The nearly perfect results across the board reinforce the model’s suitability for real-world IoT environments, where precise anomaly detection is critical for maintaining security and operational integrity.

\begin{figure}[h]
    \centering
    \includegraphics[width=3.54in]{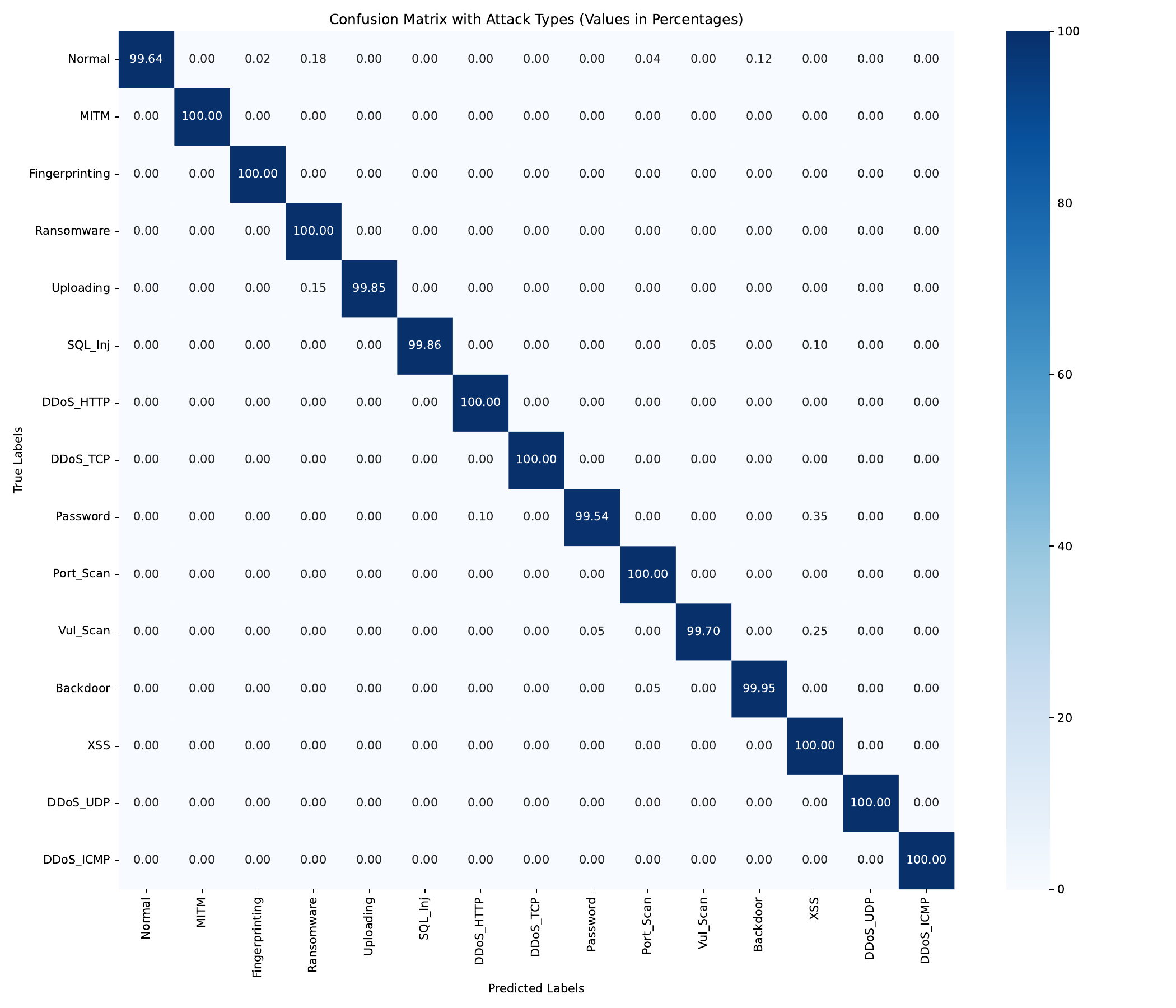}  
    \caption{Confusion matrix showcasing the classification performance of our fine-tuned BERT model for IoT anomaly detection.}
    \label{fig5}
\end{figure}

\subsection*{b) Confusion Matrix}
The fine-tuned BERT model achieves state-of-the-art performance in IoT traffic anomaly detection, as evidenced by the classification report in Table IV and the confusion matrix in Figure 6. The model attains perfect recall (100\%) across multiple attack types, including ‘MITM,’ ‘Fingerprinting,’ ‘Ransomware,’ ‘DDoS\_HTTP,’ ‘DDoS\_TCP,’ ‘Port\_Scan,’ ‘Backdoor,’ ‘XSS,’ ‘DDoS\_UDP,’ and ‘DDoS\_ICMP,’ consistently identifying these critical threats with no errors.

For ‘Normal’ traffic—the largest class with 4,985 samples—the model achieves a recall of 99.64\%, demonstrating its ability to identify almost all benign activities correctly. High recall is also observed for ‘SQL Injection’ (99.86\%) and ‘Uploading’ (99.85\%), showcasing the model’s sensitivity to subtle attack patterns. However, the lowest recall (99.54\%) occurs for the ‘Password’ class, with misclassifications primarily attributed to the ‘XSS’ category, indicating overlapping traffic characteristics between these attack types.

The model achieves weighted averages of 99.87\% for accuracy, recall, precision, and F1-score, setting a new standard for IoT anomaly detection with unparalleled precision and reliability in real-world scenarios.

\begin{table*}[ht]
    \centering
    \caption{Comparison of our fine-tuned BERT with Traditional ML and DL Models}
    \label{tab:securitybert_comparison}
    \resizebox{\textwidth}{!}{%
    \begin{tabular}{@{}llclc@{}}
        \toprule
        \textbf{AI Type} & \textbf{Authors} & \textbf{Year} & \textbf{AI Model} & \textbf{Accuracy} \\
        \midrule
        \multirow{5}{*}{Traditional ML} 
        & Ferrag et al. \cite{9751703} & 2022 & Decision Tree (DT) & 67.11\% \\
        & Ferrag et al. \cite{9751703} & 2022 & Random Forest (RF) & 80.83\% \\
        & Ferrag et al. \cite{9751703} & 2022 & Support Vector Machines (SVM) & 77.61\% \\
        & Aouedi et al. \cite{aouedi2023f} & 2023 & DT + RF / FL & 90.91\% \\
        & Zhang et al. \cite{zhang2023automatic} & 2023 & K-Nearest Neighbor (KNN) & 93.78\% \\
        & Ferrag et al. \cite{9751703} & 2022 & K-Nearest Neighbor (KNN) & 79.18\% \\
        \midrule
        \multirow{8}{*}{Deep Learning Models} 
        & Friha et al. \cite{friha20232df} & 2022 & CNN / CL / No-DP & 94.84\% \\
        & Friha et al. \cite{friha20232df} & 2023 & CNN / FL / No-DP & 93.96\% \\
        & Aljuhani et al. \cite{aljuhani2023deep} & 2023 & CSAE + ABiLSTM & 94.40\% \\
        & Friha et al. \cite{friha2022felids} & 2022 & Recurrent Neural Network (RNN) & 94.00\% \\
        & Ding et al. \cite{ding2023deepak} & 2023 & Long Short-Term Memory (LSTM) & 94.96\% \\
        & Ferrag et al. \cite{9751703} & 2022 & Deep Neural Network (DNN) & 94.67\% \\
        & Friha et al. \cite{friha2022felids} & 2022 & Deep Neural Network (DNN) & 93.00\% \\
        & E. M.d. Elias et al. \cite{de2022hybrid} & 2022 & CNN-LSTM & 97.14\% \\
        & Ferrag et al. \cite{ferrag2023generative} & 2023 & Transformer model w/o Tokenization and Embedding & 94.55\% \\
        \midrule
        \multirow{3}{*}{Large Language Model} 
        & M. A. Ferrag et al. \cite{ferrag2024revolutionizing} & 2024 & SecurityBERT with PPFLE & 98.2\% \\
        & \textit{This work} & 2024 & Fine-tuned BERT & \textbf{99.87\%} \\
        \bottomrule
    \end{tabular}}
    \caption*{CNN: Convolutional Neural Network, CL: Centralized Learning, FL: Federated Learning, DP: Differential Privacy, CSAE: Contractive Sparse AutoEncoder, ABiLSTM: Attention-based Bidirectional Long Short-Term Memory.}
\end{table*}

\begin{table}[ht]
    \centering
    \caption{Classification report of our fine-tuned BERT}
    \label{tab:classification_report}
    \resizebox{\columnwidth}{!}{%
    \begin{tabular}{@{}lcccc@{}}
        \toprule
        \textbf{Class} & \textbf{Precision} & \textbf{Recall} & \textbf{F1-Score} & \textbf{Support} \\
        \midrule
        Normal         & 100.00 & 99.64 & 99.82 & 4,985 \\
        MITM      & 100.00 & 100.00 & 100.00 & 255 \\
        Fingerprinting     & 99.49 & 100.00 & 99.75 & 197 \\
        Ransomware & 99.42 & 100.00 & 99.71 & 2,061 \\
        Uploading       & 100.00 & 99.85 & 99.93 & 2,015 \\
        SQL\_Injection      & 100.00 & 99.86 & 99.93 & 2,089 \\
        DDoS\_HTTP     & 99.91 & 100.00 & 99.95 & 2,129 \\
        DDoS\_TCP   & 100.00 & 100.00 & 100.00 & 1,961 \\
        Password      & 99.95 & 99.54 & 99.75 & 1,973 \\
        Port\_Scanning       & 99.86 & 100.00 & 99.93 & 2,128 \\
        Vul\_Scanner & 99.95 & 99.70 & 99.83 & 2,009 \\
        Backdoor            & 99.70 & 99.95 & 99.82 & 1,973 \\
        XSS     & 99.32 & 100.00 & 99.66 & 2,045 \\
        DDoS\_UDP & 100.00 & 100.00 & 100.00 & 2904 \\
        DDoS\_ICMP           & 100.00 & 100.00 & 100.00 & 2836 \\
        \midrule
        \textbf{Macro Avg} & 99.84 & 99.90 & 99.87 & 31,560 \\
        \textbf{Weighted Avg} & 99.87 & 99.87 & 99.87 & 31,560 \\
        \textbf{Accuracy} & \multicolumn{4}{c}{\textbf{ 99.87\%}} \\
        \bottomrule
    \end{tabular}}
\end{table}

\subsection*{\textbf{2) Performance Comparison}}
As detailed in Table III, the Edge-IIoTset dataset has been rigorously tested across traditional ML, DL, and LLM-based approaches. Among traditional ML models, KNN achieved the highest accuracy of 93.78\%, outperforming RF (80.83\%), SVM (77.61\%), and DT (67.11\%).

Deep learning models demonstrated significant improvements, with CNN-LSTM achieving 97.14\%, showcasing its ability to capture both spatial and temporal features. Ferrag et al.\cite{ferrag2023generative} proposed a GAN-Transformer model with 94.55\% accuracy, emphasizing challenges in tokenizing unstructured IoT traffic.

SecurityBERT introduced Privacy-Preserving Fixed-Length Encoding (PPFLE), which reached 98.2\% accuracy. Although designed for resource-constrained IoT devices, it prioritized compactness and privacy at the cost of scalability and generalization.

Our fine-tuned BERT model redefines IoT anomaly detection with an unparalleled accuracy of 99.89\%. By leveraging the pre-trained BERT model in administrative systems, our framework provides scalable, precise, and context-aware solutions for detecting complex anomalies in real-world IoT networks.

\subsection{Discussion}
The experimental findings confirm the efficacy of the framework in addressing IoT management and security challenges. The RAG with LLM module consistently enhanced response quality, achieving up to 82.95\% improvements in BLEU, ROUGE, METEOR, and BERTScore metrics. These results underscore the effectiveness of contextual augmentation in grounding LLM outputs, reducing hallucinations, and delivering precise, task-relevant responses. Resource efficiency was evident, with notable reductions in execution time and token metrics in the WC scenario, demonstrating computational optimization without compromising response quality.

The fine-tuned BERT module achieved a record-breaking accuracy of 99.87\%, demonstrating exceptional precision and recall across diverse IoT attack types. Critical threats, including ‘MITM’ and ‘DDoS\_TCP,’ were identified with near-perfect recall, affirming the model’s robustness in detecting anomalies. Minor misclassifications were limited to closely related attack types, having negligible practical impact, further highlighting the model’s reliability for real-world IoT environments.

By integrating these complementary modules, our framework delivers an efficient, scalable, and high-accuracy solution for IoT ecosystem management, establishing a benchmark for addressing complex, real-world challenges in the domain.

\section{Related Work}
\label{sec: Lit}

The integration of LLMs within IoT ecosystems has opened new avenues for addressing operational challenges and enhancing security. Existing research showcases significant advancements in task modeling, reasoning, and anomaly detection tailored to IoT environments.

\subsection{Task Reasoning and Modeling}
\label{sec: reasoning}

LLMs have extended their capabilities beyond traditional text-processing tasks to include reasoning about physical-world IoT systems. Research has shown that LLMs can process multi-modal IoT data, such as accelerometer signals, WiFi patterns, and heart rate measurements, enabling them to classify human activities, monitor environments, and analyze physical phenomena \cite{an2024iot,xu2024penetrative}. These developments highlight LLMs’ potential to interpret IoT sensor data and generate actionable insights by integrating embedded world knowledge.

To better adapt LLMs to IoT-specific tasks, researchers have introduced frameworks that incorporate sensor data preprocessing, knowledge retrieval, and advanced prompting strategies \cite{mo2024iot,xiao2024efficient}. These approaches enable LLMs to handle diverse IoT modalities, including human activity recognition, industrial monitoring, and indoor localization. Techniques such as multitasking adapter layers and structured prompting have been deployed to improve performance on complex, real-world tasks. Moreover, the creation of extensive multi-modal datasets has further enhanced LLMs' capabilities, ensuring scalability and adaptability across varied IoT applications\cite{xiao2024efficient}.

\subsection{Security} 
\label{sec: security}
In the realm of security, LLMs have shown remarkable promise in identifying and mitigating threats within IoT networks\cite{kok2024iot}. Transformer architectures have been utilized to detect advanced cybersecurity threats, such as denial-of-service attacks, reconnaissance activities, and command injection vulnerabilities\cite{hassanin2025pllm,ye2024detecting}. These models use techniques such as static taint analysis and chaining of large language models to improve accuracy and efficiency in identifying network attacks.

BERT-based models have proven particularly effective in anomaly detection for structured environments. Early applications succeeded in detecting arbitration identifier anomalies in controller area networks and malware classification tasks\cite{alkhatib2022can}. Similarly, advancements in malware classification leveraged contextual relationships to identify malicious patterns in source code\cite{rahali2021malbert}. Recent advancements have expanded these capabilities to large-scale IoT systems, treating logs and network data as sequential language-like inputs\cite{chen2022bert}. This has enabled the detection of advanced persistent threats and subtle vulnerabilities with high accuracy\cite{adjewa2024efficient,kholgh2023pac,ranade2021cybert}.

However, despite these advancements, BERT’s potential for real-time packet-level IoT traffic analysis remains underutilized\cite{ferrag2024revolutionizing}. Current efforts often emphasize privacy-preserving mechanisms or lightweight adaptations for resource-constrained devices, which restrict their effectiveness in accurately detecting nuanced anomalies in complex, dynamic IoT ecosystems\cite{ferrag2024revolutionizing,wang2024lightweight}. Unlocking BERT’s full contextual understanding offers a compelling opportunity to enhance anomaly detection and ensure robust IoT network security.

Despite advancements in applying LLMs to IoT environments, existing approaches are limited in scalability and adaptability to IoT ecosystems' diverse and dynamic nature. Additionally, no frameworks comprehensively address administrative functionalities—such as device management, troubleshooting, device setup, maintenance, and security and privacy—by leveraging contextual knowledge from IoT manuals and related documentation while incorporating IoT traffic analysis for anomaly detection to support administrative purposes. This gap highlights the pressing need for an integrated solution that unifies operational management with robust anomaly detection and is designed to meet the complexities of IoT networks.

\section{Conclusion} 
This paper proposes a framework that integrates a context-aware generation module and a fine-tuned BERT model, addressing critical challenges in IoT device management and network security. The RAG component bridges the gap between generic LLMs and IoT-specific demands, enabling precise, contextually grounded responses across five administrative use cases: device management, maintenance, security and privacy, troubleshooting, and device setup. Meanwhile, the fine-tuned BERT model excels in anomaly detection, achieving state-of-the-art accuracy in identifying diverse attack types and ensuring robust security for IoT ecosystems.

Our experimental results demonstrate that contextual augmentation significantly enhances LLM performance, reducing hallucinations and increasing the reliability of responses, while the anomaly detection module ensures actionable insights for cybersecurity. By combining these strengths, our framework delivers a scalable, efficient, and practical solution to the growing complexity of IoT environments, empowering administrators to manage and secure devices with unprecedented accuracy. This work lays a solid foundation for advancing AI-driven IoT systems and offers a path forward for future research in intelligent, secure IoT ecosystem management.

\bibliography{shepherd}

\bibliographystyle{IEEEtran}

\end{document}